\documentclass[useAMS,usenatbib,usegraphicx]{mn2e}

\usepackage[a4paper, centering, totalwidth=520pt, totalheight=700pt]{geometry}

\begin{document}

\title{Dissipationless collapse of a set of $N$ massive particles} 
\author[F.Roy and J.Perez]{Fabrice Roy$^{(1)}$\thanks{
roy@ensta.fr} and J\'{e}r\^{o}me Perez$^{(1,2)}$\thanks{
perez@ensta.fr} \\
$^{(1)}$\'Ecole Nationale Sup\'{e}rieure de Techniques Avanc\'{e}es,
Unit\'{e} de Math\'{e}matiques Appliqu\'{e}es, 32\ Bd Victor, 75015 Paris, France\\
$^{(2)}$Laboratoire de l'Univers et de ses TH\'{e}ories, Observatoire de
Paris-Meudon, 5 place Jules Janssen, 92350 Meudon, France}
\maketitle

\begin{abstract}
The formation of self-gravitating systems is studied by simulating the
collapse of a set of N particles which are generated from several
distribution functions. We first establish that the results of such
simulations depend on N for small values of N. We complete a previous work by Aguilar \& Merritt concerning
the morphological segregation between spherical and elliptical equilibria. We find and interpret two new segregations: one concerns the equilibrium  core size and the other the equilibrium temperature. All these features are used to explain some of the global properties of self-gravitating objects: origin of globular clusters and central black hole or shape of elliptical galaxies.
\end{abstract}

\begin{keywords}
methods: numerical, N-Body simulations -- galaxies: formation -- globular clusters: general.
\end{keywords}

\section{Introduction}

It is intuitive that the gravitational collapse of a set of $N$ masses is
directly related to the formation of astrophysical structures like globular
clusters or elliptical galaxies (the presence of gas may complicate the pure
gravitational $N$-body problem for spiral galaxies). From an analytical
point of view, this problem is very difficult. When $N$ is much larger than
2, direct approach is intractable, and since Poincar\'{e} results of non
analyticity, exact solutions may be unobtainable. In the context of
statistical physics, the situation is more favorable and, in a dissipationless
approximation\footnote{
The dissipationless hypothesis is widely accepted in the context of
gravitational $N$-body problem because the ratio of the two-body relaxation
time over the dynamical time is of the order of $N$. For a system composed
of more than $\sim 10^{4}$ massive particles a study during a few hundreds
dynamical times can really be considered as dissipationless, the unique
source of dissipation being two-body encounters.}, leads to the
Collisionless Boltzmann Equation (hereafter denoted by CBE) 
\begin{equation}
\frac{\partial f}{\partial t}+\mathbf{p}.\frac{\partial f}{\partial \mathbf{
r }}+m\frac{\partial \psi }{\partial \mathbf{r}}.\frac{\partial f}{\partial 
\mathbf{p}}=0  \label{cbe}
\end{equation}
where $f=f\left( \mathbf{r},\mathbf{p},t\right) $ and $\psi =\psi \left( 
\mathbf{r},t\right) $ are respectively the distribution function of the
system with respect to the canonically conjugated $\left( \mathbf{r},\mathbf{p}\right)$
phase space variables and the mean field gravitational potential. As noted
initially by \citet{henon}, this formalism holds for such systems if and
only if we consider $N$ identical point masses equal to $m$. This problem
splits naturally into two related parts: the time dependent regime and the
stationary state. We can reasonably think that these two problems are not
completely understood. The transient time dependent regime was investigated
mainly considering self-similar solutions (\citet{LBE}, \citet{HW}, 
\citet{bouquet} and \citet{lance}). These studies conclude that power law
solutions can exist for the spatial dependence of the gravitational
potential (with various powers). Nevertheless, there is no study which indicates
clearly that the time dependence of the solutions disappears in a few
dynamical times, giving a well defined equilibrium-like state. On the other
hand, applying Jeans theorem (e.g. \citet{BT87} hereafter BT87, p. 220), it is quite easy to
find a stationary solution. For example, every positive and integrable
function of the mean field energy per mass unit $E$ is a potential
equilibrium distribution function for a spherical isotropic system. Several
approaches are possible to choose the equilibrium distribution function.
Thermodynamics (Violent Relaxation paradigm: \citet{LB67}, \citet{C}, \cite
{J}) indicate that isothermal spheres or polytropic systems are good
candidates. Stability analyses can be split into two categories. In the
CBE context (see \cite{PA} for a review), it is well known that spherical
systems (with decreasing spatial density) are generally stable except in the
case where a large radial anisotropy is present in the velocity space. This
is the Radial Orbit Instability, hereafter denoted by ROI (see \cite{PA}, and \cite
{PAA98} for a detailed analytic and numeric study of these phenomena) which
leads to a bar-like equilibrium state in a few dynamical times. In the
context of thermodynamics of self-gravitating systems, in a pioneering work
by \cite{ANT62}, it was shown that an important density contrast leads to the
collapse of the core of system (see \cite{CH2} for details). \newline
In all these studies there is no definitive conclusion, and the choice of
the equilibrium distribution remains unclear. Introducing observations and
taking into account analytical constraints, several models are possible:
chronologically, we can cite (see for example BT87, p. 223-239) the
Plummer model (or other polytropic models), de Vaucouleurs $r^{1/4}$ law,
King and isochrone H\'enon model or more recently the very simple but
interesting Hernquist model (\cite{Hernquist90}) for spherical isotropic
systems. In the anisotropic case, Ossipkov-Merritt or generalized polytropes
can be considered. Finally for non spherical systems, there also exists some
models reviewed in BT87 (p. 245-266). Considering this wide variety of
possibilities, one can try to make accurate numerical simulations to clarify
the situation. Surprisingly, such a program has not been completely carried on.
In a pioneering work, \cite{vA82} remarked that the dissipationless collapse
of a clumpy cloud of $N$ equal masses could lead to a final stationary state
that is quite similar to elliptical galaxies. This kind of study was
reconsidered in an important work by \cite{AM90}, with more details and a crucial
 remark concerning the correlation between the final shape
(spherical or oblate) and the virial ratio of the initial state. These
authors explain this feature invoking ROI. Some more recent studies (\cite{CH92}, 
\cite{BCM99} and \cite{STH}) concentrate on some particularities of the preceding works.
Finally, two works (\cite{DCdCR02} and \cite{CM95})
develop new ideas considering the influence of the Hubble flow on the
collapse. However, the problem is only partially depicted. \newline
The aim of this paper is to analyse the dissipationless collapse of a large
set of N Body systems with a very wide variety of `realistic' initial
conditions. As we will see, the small number of particles involved, the
numerical technique or the specificity of the previous works did not allow
their authors to reach a sufficiently precise conclusion. The layout of
this paper is as follows. In section 2 we describe in detail the numerical
procedures used in our experiments. Section 3 describes the results we have obtained. 
These results are then interpreted in section 4, where some conclusions and 
perspectives are also proposed.

\section{Numerical procedures}

\subsection{Dynamics}

The Treecode used to perform our simulations is a modified version of the 
\cite{BH86} Treecode, parallelised by D. Pfenniger using the MPI library. We
implemented some computations of observables and adapted the code to suit our
specific problems. The main features of this code are a hierarchical $
O(N\log (N))$ algorithm for the gravitational force calculation and a
leap-frog algorithm for the time integration. We introduced an adaptative
time step, based on a very simple physical consideration. The time step is
equal to a fraction $n_{ts}$ of the instantaneous dynamical time $T_d$
(\footnote{The fraction $n_{ts}$ is adapted to the virial parameter $\eta$ and ranges roughly from $n_{ts}=300$ when $\eta=90$ to $n_{ts}=5000$ when $\eta=08$. The dynamical time we used is given by
\[
T_d = \frac
{\displaystyle{\sum_{i=1}^{N} \sqrt{ x_i^2+ y_i^2+ z_i^2}}}
{\displaystyle{\sum_{i=1}^{N} \sqrt{vx_i^2+vy_i^2+vz_i^2}}}
\]
})
, i.e. 
$\Delta t = T_d / n_{ts}$
. The simulations were run on a Beowulf cluster (25
dual CPU processors whose speed ranges from 400MHz to 1GHz).

\subsection{Initial Conditions}

The initial virial ratio is an important parameter in our simulations. The
following method was adopted to set the virial ratio to the value $
V_{initial}$. Positions $\mathbf{r}_{i}$ and velocities $\mathbf{v}_{i}$ are generated. We
can then compute 
\begin{equation}
V_{p}=\frac{2K}{U}\ 
\end{equation}
where 
\begin{equation}
K=\sum_{i=1}^{N}\frac{1}{2}m_{i}\mathbf{v}_{i}^{2}
\end{equation}
and 
\begin{equation}
U=- \frac{G}{2}\sum_{i\neq j}^{N} \frac{m_{i}m_{j}} {\left( max((\mathbf{r}
_{i}-\mathbf{r}_{j})^{2},\epsilon^{2})\right)^{1/2} }  \label{defpot} \,\,\, .
\end{equation}
In this relation $\epsilon$ is a softening parameter whose value is
discussed in section \ref{soft}. As the potential energy depends only on the
positions, we obtain a system with a virial ratio equal to $V_{initial}$
just by multiplying all the particle velocities by the factor $\left(
V_{initial}/V_{p}\right) ^{1/2}$.\ For convenience we define 
\begin{equation}
\eta =\left| V_{initial}\right| \times 10^{2}
\end{equation}

\subsubsection{Homogeneous density distribution (H$_{\eta }$)}

As we study large $N$-body systems, we can produce a homogeneous density by
generating positions randomly. These systems are also isotropic. We produce
the isotropic velocity distribution by generating velocities randomly.

\subsubsection{Clumpy density distribution (C$_{\eta }^{n}$)}

A type of inhomogeneous systems is made of systems with a clumpy density
distribution. We first generate $n$ small homogeneous spherical systems with
radius $R_{g}$. Centers of these subsystems are uniformly distributed in the
system. The empty space is then filled using a
homogeneous density distribution. In the initial state, each clump contains about 1\% of the total mass of the system and have a radius which represents $5\%$ of the initial radius of the whole system.
These systems are  isotropic.

\subsubsection{Power law $r^{-\alpha }$ density distribution (P$_{\eta
}^{\alpha }$)}

We first generate the $\varphi$ and $z$ cylindrical
coordinates using two uniform random numbers $u_{1}$ and $u_{2}$: 
\begin{equation}
\left( z,\varphi \right) =\left( 2u_{1}-1,2\pi u_{2}\right)\;\;.
\end{equation}
Using the inverse transformation method, if
\begin{equation}
r=RF^{-1}\left( u\right) \quad \mbox{with\quad }F\left( r\right) =\frac{1}{S}
\int_{\iota}^{r}x^{2-\alpha}dx
\end{equation}
where $R$ is the radius of the system, $u$ is a uniform random number, $
\iota \ll 1$ and 
\begin{equation}
S=\int_{0}^{1}x^{2-\alpha}dx \;\; ,
\end{equation}
then the probability density of $r$ is proportional to $r^{2-\alpha}$, and the mass density $\rho$ is proportional to  $r^{-\alpha }$. Finally, one gets 
\begin{equation}
\mathbf{r}=\left[ 
\begin{array}{c}
r\sqrt{1-z^{2}}\cos \varphi \\ 
r\sqrt{1-z^{2}}\sin \varphi \\ 
rz
\end{array}
\right]
\end{equation}
These systems are isotropic.

\subsubsection{Gaussian velocity distribution(G$_{\eta }^{\sigma }$)}
Most of the systems we use have a uniform velocity distribution. But we have also
performed simulations with systems presenting a gaussian initial velocity distribution.
These systems are isotropic, but the x-, y- and z-components of the velocity
are generated following a gaussian distribution. Using a standard method
we generate two uniform random numbers $u_{1}$ and $u_{2}$, and set 
\begin{equation}
v_{i}=\sqrt{-2 \sigma^{2}\ln u_{1}}\cos (2\pi \sigma^{2}u_{2})\qquad i=x,y,z
\end{equation}
where $\sigma$ is the gaussian standard deviation.

\subsubsection{Global rotation (R$_{\eta }^{f}$) \label{secrotation}}

Some of our initial systems are homogeneous systems with a global rotation
around the $Z$-axis. The method we choose to generate such initial conditions is the
following. We create a homogeneous and isotropic system (an H-type system).
We then compute the average velocity of the particles. 
\begin{equation}
\bar{v}=\frac{1}{N}\sum_{i=1}^{N}{\Vert \mathbf{v}_{i}\Vert }
\end{equation}
We project the velocities on a spherical referential, and add a fraction of $
\bar{v}$ to $v_{\phi }$ with regard to the position of the particle. We set 
\begin{equation}
v_{i,\phi }=v_{i,\phi }+f\frac{\rho _{i}\bar{v}}{R}
\end{equation}
where $f$ is a parameter of the initial condition, $\rho _{i}$ is the
distance from the particle to the $Z$-axis and $R$ the radius of the system.
The amount of rotation induced by this method can be evaluated through
the ratio: 
\begin{eqnarray}
\mu = K_{rot} / K,
\end{eqnarray}
\noindent where $K$ is the total kinetic energy defined above, whereas $
K_{rot} $ is the rotation kinetic energy defined by \cite{navaro}: 
\begin{equation}
K_{rot}=\frac{1}{2}\sum_{i=1}^{N} m_i
 \frac {
 (\mathbf{J}_i \cdot \hat{\mathbf{J}}_{tot})^2} 
 {[\mathbf{r}_{i}^{2}-(\mathbf{r}_i \cdot \hat{\mathbf{J}}_{tot})^2]}  \label{krot}
\end{equation}
Above, $\mathbf{J}_i$ is the specific angular momentum of particle
i, $\hat{\mathbf{J}}_{tot}$ is a unit vector in the direction of the
total angular momentum of the system. In order to exclude counter-rotating
particles, the sum in equation (\ref{krot}) is actually carried out only
over those particles which verify the condition $(\mathbf{J}_i \cdot 
\hat{\mathbf{J}}_{tot}) > 0$.

\subsubsection{Power-law initial mass function(M$_{\eta }^{k}$)}

Almost all the simulations we made assume particles with equal masses. However, we
have created some initial systems with a power-law mass function, like
\begin{equation}
n(M)=\alpha M^{\beta }
\end{equation}
The number of particles whose mass is $M\leq m\leq M+dM$ is $n(M)dM$. In
some models, the value of $\alpha $ and $\beta $ depends on the 
range of mass that is considered. We have used several types of mass functions, among them the
initial mass function given by \cite{Kroupa} $(k=I)$, the one given by \cite
{Salpeter} $(k=II)$ and an $M^{-1}$ mass function $\left( k=III\right)$. In
order to generate masses following these functions, we first calculate $
\alpha _{k}$ to produce a continuous function. We then can calculate the
number of particles whose mass is between $M$ and $M+dM$. We
generate $n(M)$ masses 
\begin{equation}
m_{i}=M+udM\hspace{2cm}1\leq i\leq n(M)
\end{equation}
where $0\leq u\leq 1$ is a uniform random number. In the initial state, these systems have a
homogeneous number density, a quasi homogeneous mass density and they are isotropic.

\subsubsection{Nomenclature}

We indicate below the whole set of our non rotating initial conditions.

\begin{itemize}
\item Homogeneous H$_{\eta }$ models: H$_{88}$, H$_{79}$, H$_{60}$, H$_{50}$, H$_{40}$, H$_{30}$, H$_{20}$, H$_{15}$ and H$_{10}$

\item Clumpy C$_{\eta }^{n}$ models: C$_{67}^{20}$, C$_{65}^{20}$, C$_{61}^{20}$, C$_{48}^{20}$, C$_{39}^{20}$, C$_{29}^{20}$, C$_{14}^{20}$, C$_{10}^{20}$, C$_{07}^{20}$ and C$_{10}^{03}$

\item Power Law P$_{\eta }^{\alpha }$ models: P$_{50}^{2.0}$, P$_{09}^{2.0}$, P$_{50}^{1.0}$, P$_{50}^{0.5}$, P$_{10}^{1.0}$, P$_{08}^{1.5}$ and P$_{40}^{1.5}$

\item Gaussian velocity profiles G$_{\eta }^{\sigma }$ models: G$_{50}^{1}$, G$_{50}^{2}$, G$_{50}^{3}$, G$_{12}^{4}$ and G$_{50}^{5}$

\item Mass spectra M$_{\eta }^{k}$ models: M$_{50}^{I}$, M$_{50}^{II}$, M$_{51}^{III}$, M$_{35}^{I}$, M$_{25}^{II}$, M$_{15}^{III}$ and M$_{07}^{I}$
\end{itemize}

For all these models we ran the numerical simulations with 30 000 particles (see \S\ \ref{number})

\subsection{Observables}

\subsubsection{Units}

Our units are not the commonly used ones (see \cite{heggie}). We did not
set the total energy $E $ of the system to $-0.25$ because we wanted to prescribe instead
the initial virial ratio $V_{initial}$, the size $R$ of the system and its
mass $M$. We thus have $M=1 $, $R=10$ and $G=1$, and values of $V_{initial}$
and $E$ depending on the simulation. We can link the units we have used with
more standard ones. We have chosen the following relationships
between our units of length and mass and common astrophysical ones: 
\begin{equation}
M=10^{6}M_{\sun }\mbox{ and }R=10 \; pc
\end{equation}
Our unit of time $u_t$ is given by: 
\begin{equation}
1u_{t}=\sqrt{\frac{R_{c}^{3}G_{s}M_{s}}{R_{s}^{3}G_{c}M_{c}}}\approx
4.72\,\,10^{11}\;s=1.50\,\,10^{4}\;yr
\end{equation}
where variables $X_{s}$ are expressed in our simulation units and
variables $X_{c}$ in standard units.

\subsubsection{Potential softening and energy conservation\label{soft}}

The non conservation of the energy during the numerical evolution has three
main sources.\newline
The softening parameter $\varepsilon $ introduced in the potential calculus
(cf. equation \ref{defpot}) is an obvious one. This parameter
introduces a lower cutoff $\Lambda $ in the resolution of length in the simulations.
Following \cite{BH89}, structural details up to scale $\Lambda \la 
10\varepsilon $ are sensitive to the value of $\varepsilon $. Moreover, in
order to be compatible with the collisionless hypothesis, the softening
parameter must be greater than the scale where important collisions can
occur. Still following \cite{BH89}, this causes 
\begin{equation}
\frac{\varepsilon }{10}\ga \frac{G\left\langle m\right\rangle }{\left\langle
v^{2}\right\rangle }
\end{equation}
In our collapse simulation with $3\cdot10^4$ particles, this results in $
\varepsilon \eta \ga 2/3$. The discretization of time integration
introduces inevitably another source of energy non conservation, particularly during the collapse. The force computation also generates errors. The choice of the
opening angle $\Theta$, which governs the accuracy of the force calculation
of the Treecode, is a compromise between speed and accuracy. For all these
reasons, we have adopted $\varepsilon =0.1$. This choice imposes $\eta \ga 6$
(for $30\cdot10^4$ particles). This trade-off allowed to perform simulations
with less than $1\%$ energy variation without requiring too much computing time.
For each of our experiments, the total CPU time ranges between 3 to 24 hours
for $3000\;u_t$ and $3\cdot10^4$ particles. The total agregated CPU time of all our
collapse experiments is approximately 6 months.\newline
We have tested two other values of the softening parameter ($\varepsilon =
0.03$ and $\varepsilon = 0.3$) for several typical simulations. These tests
did not reveal significant variations of the computed observables.

\subsubsection{Spatial indicators}

As indicators of the geometry of the system, we computed axial ratios, radii
containing $10\%$ ($R_{10}$), $50\%$ ($R_{50}$) and $90\%$ ($R_{90}$) of the
mass, density profile $\rho \left( r\right) $ and equilibrium core radius.
The axial ratios are computed with the eigenvalues $\lambda _{1}$, $\lambda
_{2}$ and $\lambda _{3}$ of the (3x3) inertia matrix $I$, where $\lambda
_{3}\leq \lambda _{2}\leq \lambda _{1}$ and, if the position of the particle 
$i$ is $\mathbf{r}_{i}=\left(x_{1,i}\,;x_{2,i}\,;\,x_{3,i}\right) $ 
\begin{equation}
\left\{ 
\begin{array}{lll}
I_{\mu \nu }=-\sum\limits_{i=1}^{N}m_{i}\,x_{\mu ,i}\,x_{\nu ,i} &  & 
\mbox{for }\mu \neq \nu =1,2,3 \\ 
&  &  \\ 
I_{\mu \mu }=\sum\limits_{i=1}^{N}m_{i}(r_{i}^{2}-x_{\mu ,i}^{2}) &  & 
\mbox{for }\mu =1,2,3
\end{array}
\right.
\end{equation}
The axial ratios $a_{1}$ and $a_{2}$ are given by $a_{2}=\lambda
_{1}/\lambda _{2}\geq 1.0$ and $a_{1}=\lambda _{3}/\lambda _{2}\leq 1.0$.
\newline
The density profile $\rho $, which depends only on the radius $r$, together with the 
$R_{\delta}$ ($\delta=10,50,90$) have a physical meaning only for spherical or
nearly spherical systems. For all the spatial indicators computations we
have only considered particles whose distance to the center of mass of the
system is less than $6\times R_{50}$ of the system. This assumption excludes
particles which are inevitably `ejected' during the collapse\footnote{The number of excluded particles ranges from 0\% to 30\% of the total number of particles, depending mostly on $\eta$. For example, the number of excluded particles is 0\% for H$_{80}$, 3\% for C$^{20}_{67}$, 5\% for H$_{50}$, 22\% for C$^{20}_{10}$ and 31\% for H$_{10}$.}.\newline
After the collapse a core-halo structure forms in the system. In order to
measure the radius of the core, we have computed the density radius as
defined by Casertano and Hut (see \cite{CH85}). The density radius is a good
estimator of the theoretical and observational core radius.\newline
We have also computed the radial density of the system. The density is
computed by dividing the system into spherical bins and by calculating the
total mass in each bin.

\subsubsection{Statistical indicators}

When the system has reached an equilibrium state, we compute the temperature
of the system 
\begin{equation}
T=\frac{2\left\langle K\right\rangle }{3Nk_{B}}  \label{temperature}
\end{equation}
where $K$ is the kinetic energy of the system, $k_{B}$ is the Boltzmann
constant (which is set to 1) and the notation $\left\langle
A\right\rangle $ denotes the mean value of the observable $A$, defined by 
\begin{equation}
\left\langle A\right\rangle =\frac{1}{N}\sum_{i=1}^{N}A_{i}
\end{equation}
In order to characterise the system in the velocity space we have computed
the function 
\begin{equation}
\kappa \left( r\right) =\frac{2\left\langle v_{i,rad}^{2}\right\rangle _{r
\leq r_i < r+dr}}{\left\langle v_{i,tan}^{2}\right\rangle _{r \leq r_i <
r+dr}}  \label{kfunction}
\end{equation}
where $v_{i,rad}$ is the radial velocity of the i\textsuperscript{th}
particle, and $v_{i,tan}$ its tangential velocity. For spherical and
isotropic systems ($a_{1}\simeq a_{2}\simeq 1$ and $\kappa \left( r\right)
\simeq 1$), we have fitted the density by \\
1- a polytropic law 
\begin{equation}
\rho =\rho _{0}\psi ^{\gamma }  \label{df1}
\end{equation}
which corresponds to a distribution function 
\begin{equation}
f\left( E\right) \propto E^{\gamma -3/2}  \label{df2}
\end{equation}
2- an isothermal sphere law 
\begin{equation}
\rho =\rho _{1}e^{\psi /s^{2}}  \label{df3}
\end{equation}
which corresponds to a distribution function 
\begin{equation}
f\left( E\right) =\frac{\rho _{1}}{\left( 2\pi s^{2}\right) ^{3/2}}e^{\frac{E
}{s^{2}}}  \label{df4}
\end{equation}
Using the least square method in the $\ln (\rho )-\ln (\psi )$ plane we get $
\left( \gamma ,\ln \left( \rho _{0}\right) \right) $ and $\left( s^{2},\ln
\left( \rho _{1}\right) \right) $.

\section{Description of the results}

We have only studied systems with an initial virial ratio corresponding to $
\eta \in \lbrack 7,88]$. In such systems, the initial velocity dispersion
cannot balance the gravitational field. This produces a collapse. After this
collapse, in all our simulations the system reaches an equilibrium
state characterised by a temporal mean value of the virial ratio equal to $
-1 $, i.e $\eta =100$, and stationary physical observables. These quantities
(defined in section 2) are presented in a table of results in the appendix
of this paper. The following results will be discussed and interpreted in section \ref{intcon}.

\subsection{Relevant number of particles\label{number}}

In all previous works on this subject (\cite{vA82}, \cite{AM90}, \cite
{CH92} and \cite{BCM99}), the authors did not really consider the influence
of the number of particles on their results. In the first two and more
general works, this number is rather small (not more than a few thousands in
the largest simulations). The two other studies are more specific and use
typically $10^{4}$ and, in a few reference cases, $2.10^{4}$ particles. 
\begin{figure} 
\begin{center}
\includegraphics[width=84mm]{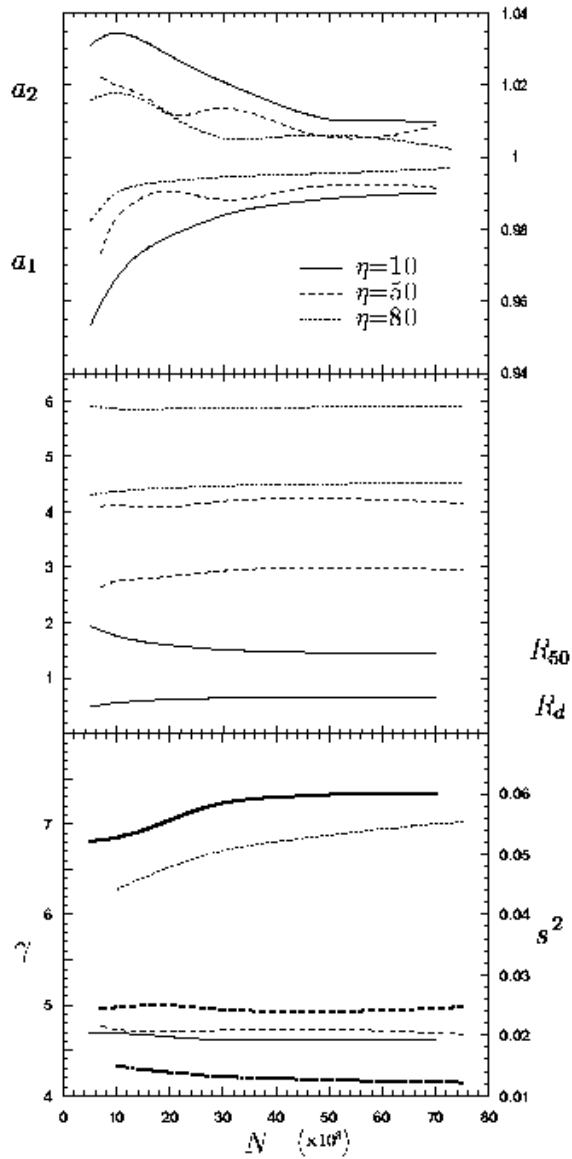}
\end{center}
\caption{Influence of the number of particles on the physical observables of
a collapsing system. Axial ratios are on the top panel, radius containing
50\% of the total mass and density radius are on the middle pannel and the
best $s^{2}$ and $\gamma $ fit for respectively isothermal and polytropic
distribution function are on the bottom panel. All cases are
initially homogeneous with $\eta =10$ (solid line), $\eta =50$ (dotted line)
and $\eta =80$ (dashed line). The number $N$ of particles used is in units
of 10$^{3}$.}
\label{nombre}
\end{figure}
In order to test the influence of the number of particles on the final
results, we have computed several physical observables of some collapsing
systems with various numbers of particles. The results are presented in
Figure \ref{nombre}. In order to check the influence of $N$ in the whole
phase space, we have studied positions and velocities related observables: $
a_1$, $a_2$, $R_{10}$, $R_{50}$ and $R_{90}$ and parameters of isothermal
and polytropic fit models namely $\gamma$ and $s^2$. Moreover, in order to
be model-independent, we have studied three representative initial
conditions: H$_{80}$, H$_{50}$ and H$_{10}$, i.e. respectively initially
hot, warm and cold systems. The number of particles used in each case ranges
from $10^{2}$ to $10^{5}.$ We can see in Figure \ref{nombre} that some
observables are N-dependent when $N<3.10^4$. In particular, $R_{50}$ and $
R_{d}$ present a monotonic variation larger than $50\%$ when $N$ varies
from $10^{2}$ to $10^{5}$ and the ellipticity of the final state is
overestimated for small values of $N$. As a conclusion of this preliminary
study, we claim that the relevant number of particles for collapse
simulations is $N\geq 3.10^{4}$. As all simulations have been completed with
a total energy loss smaller than 1\%, we state moreover that this result is
independent of the numerical scheme used (Treecode or Direct N-Body).
As a consequence, the simulations presented hereafter have been performed
using $N=3.10^{4}$ particles.

\subsection{Morphological segregation\label{morphseg}}

\begin{figure} 
\begin{center}
\includegraphics[width=84mm]{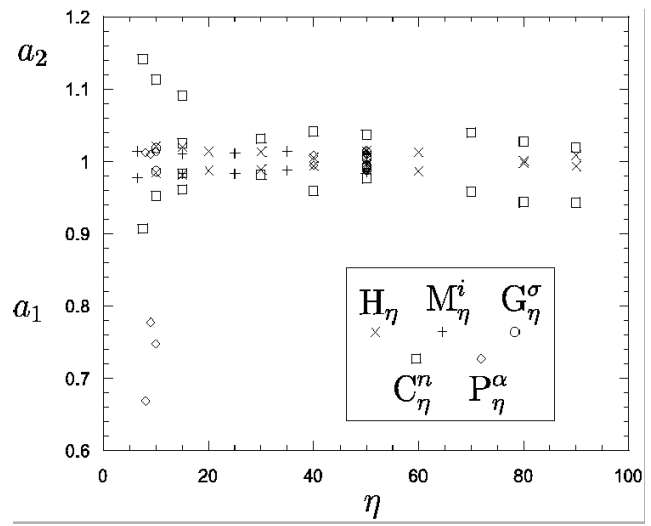}
\end{center}
\caption{Axial ratios of equilibrium states reached from Homogeneous,
Clumpy, Gaussian velocity dispersion, Power law and Mass spectrum initial
conditions.}
\label{axialratio}
\end{figure}

An important study by \cite{AM90} shows that, in the case of an initial density
profile $\rho \propto r^{-1}$, the shape of the virialized state depends on $
\eta$: a very small $\eta$ leads to a flattened equilibrium state, when a
more quiet collapse produces a spherical one. Our investigations concern a
wide range of different initial conditions (homogeneous, clumpy,...) and
show that the influence of $\eta$ depends on the properties of the
initial system\footnote{
The particular case of rotating initial conditions is discussed in a special
section.}. Figure \ref{axialratio} shows the axial ratios of the equilibrium
state reached by our simulations. In fact, only a few simulations produced a
final state with an ellipticity greater than E1. Every homogeneous initial
condition (i.e. H$_{\eta}$, G$^{\sigma}_{\eta}$ and M$^{k}_{\eta}$) resulted in
a spherical equilibrium state independently of the values of $\eta$ we
tested. Cold clumpy systems have a weakly flattened equilibrium state. The only final systems with an ellipticity
significantly greater than E1 are those produced by the collapse of cold P$^{\alpha}_{\eta}$.\newline
Previous studies invoked ROI to explain this morphological segregation. However, it seems that ROI requires inhomogeneities near the center to be triggered. 
\begin{figure}
\begin{center}
\includegraphics[width=84mm]{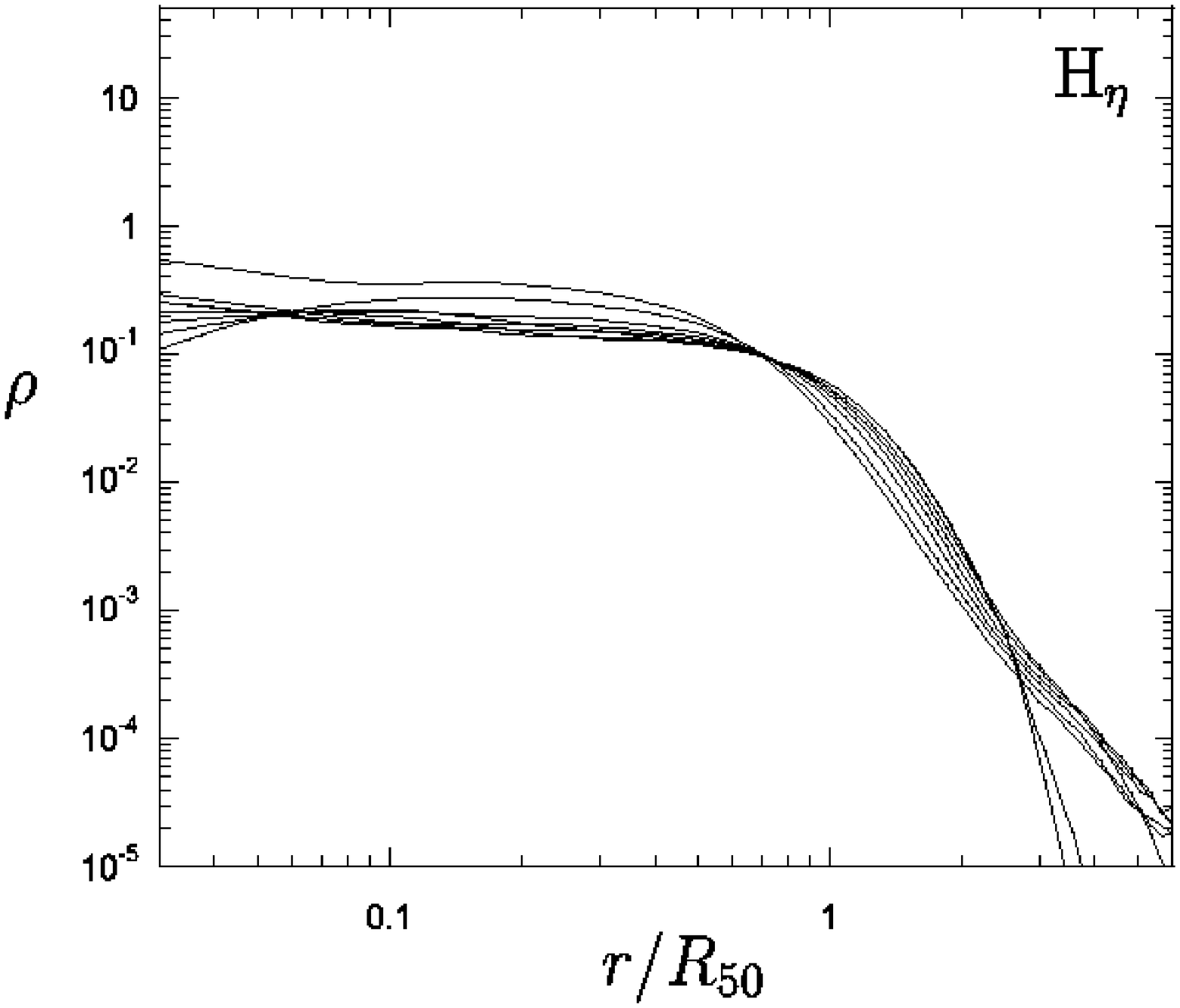}
\end{center}
\caption{Density profile for H$_{\eta }$ models plotted in units of $R_{50}$
. }
\label{densh}
\end{figure}
\begin{figure} 
\begin{center}
\includegraphics[width=84mm]{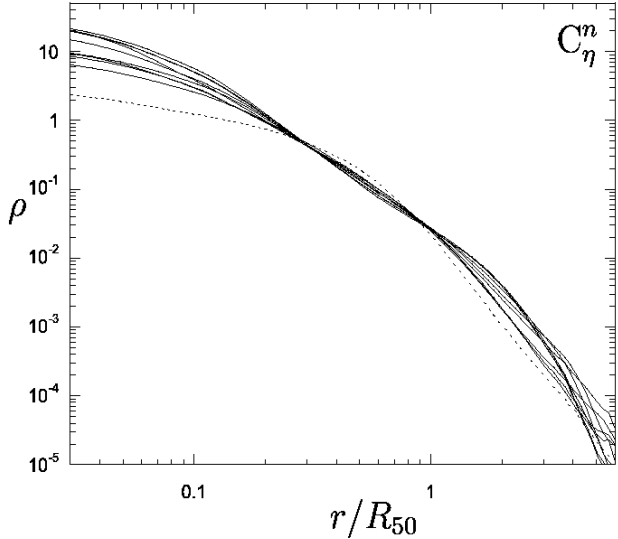}
\end{center}
\caption{Density profile for C$_{\eta }^{n}$ models plotted in units of $
R_{50}$. The dashed line corresponds to the C$_{10 }^{03}$ model}
\label{densc}
\end{figure}
\begin{figure} 
\begin{center}
\includegraphics[width=84mm]{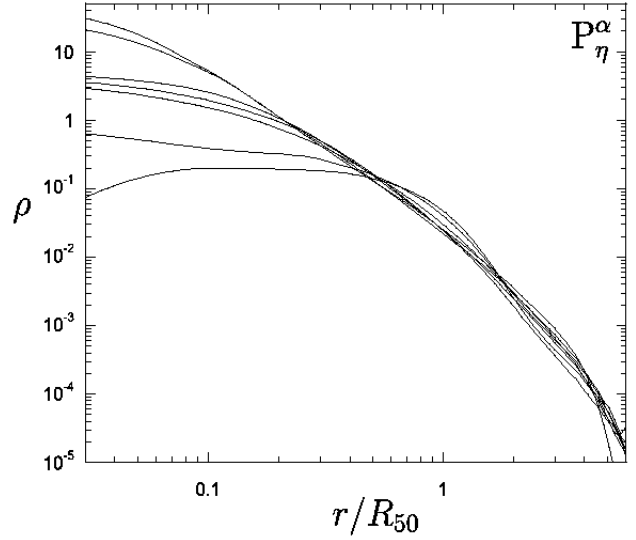}
\end{center}
\caption{Density profile for P$_{\eta}^{\alpha}$ models plotted in units of $
R_{50}$.}
\label{densp}
\end{figure}
\begin{figure}
\begin{center}
\includegraphics[width=84mm]{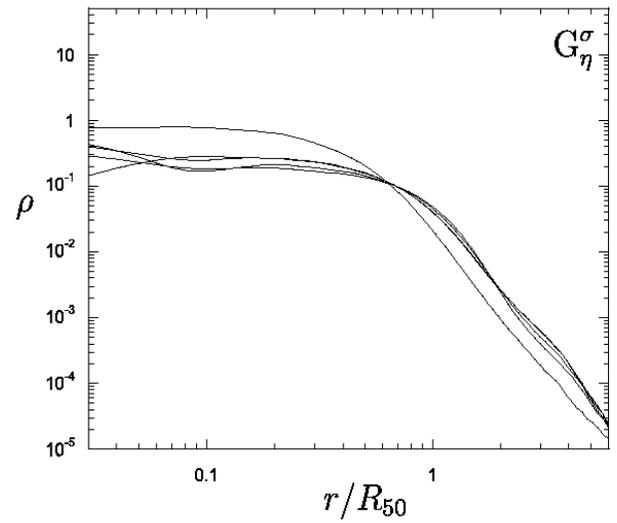}
\end{center}
\caption{Density profile for G$_{\eta }^{\sigma }$ models plotted in units
of $R_{50}$.}
\label{densg}
\end{figure}
\begin{figure}
\begin{center}
\includegraphics[width=84mm]{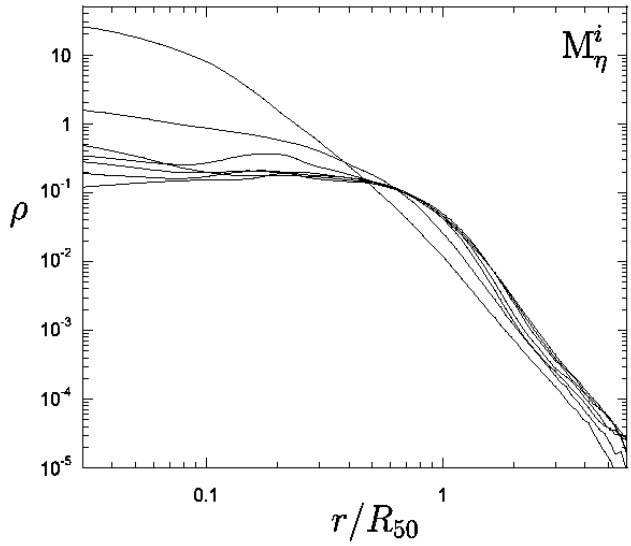}
\end{center}
\caption{Density profile for M$_{\eta }^{k}$ models plotted in units of $R_{50}$.}
\label{densm}
\end{figure}

\subsection{Characteristic size segregation\label{sizeseg}}

\begin{figure}
\begin{center}
\includegraphics[width=84mm]{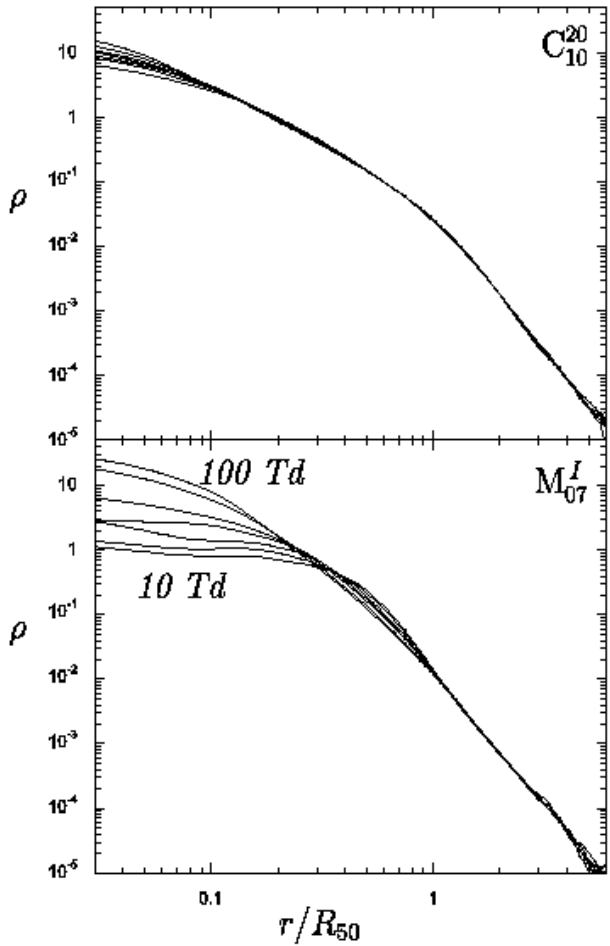}
\end{center}
\caption{Comparison between the evolution of the mass density with respect to time for C$^{20}_{10}$ (top-panel) and M$^{I}_{07}$. For each case, plotted times are 10,20,30,40,50,75 and 100 T$_d$}
\label{comparplot}
\end{figure}

In addition to the morphological segregation, presented in the previous
section, we discovered a finer phenomenon.\newline
In the Figures \ref{densh}, \ref{densc}, \ref{densp}, \ref{densg} and \ref
{densm}, we have plotted the mass density of all equilibrium states
produced by the collapse of our initial conditions as a function of the
ratio $r/R_{50}$. These plots represent the density at the end of our simulations (after about 100 crossing times). These functions do not significantly evolve after the collapse except for M$_{07}^{I}$.
For this special case, a comparative plot is the subject of Figure \ref{comparplot}.\\
All equilibrium states we obtain clearly fall into two categories:
\begin{itemize}
\item  Flat Core Systems \newline
All these systems present a core halo structure, i.e. a large central region
with a constant density and a steep envelope. These systems are typically
such that $\ln \left( R_{50}/R_{d}\right) <0.5$ and $\ln \left(
R_{10}/R_{d}\right) <-0.05$.

\item  Small Core Systems \newline
For such systems, the central density is two order of magnitude larger than
for Flat Core systems. There is no central plateau and the density falls
down regularly outward. These systems are typically such that $\ln \left(
R_{50}/R_{d}\right) >0.7$ and $\ln \left( R_{10}/R_{d}\right) >0.1$.
\end{itemize}

\begin{figure} 
\begin{center}
\includegraphics[width=84mm]{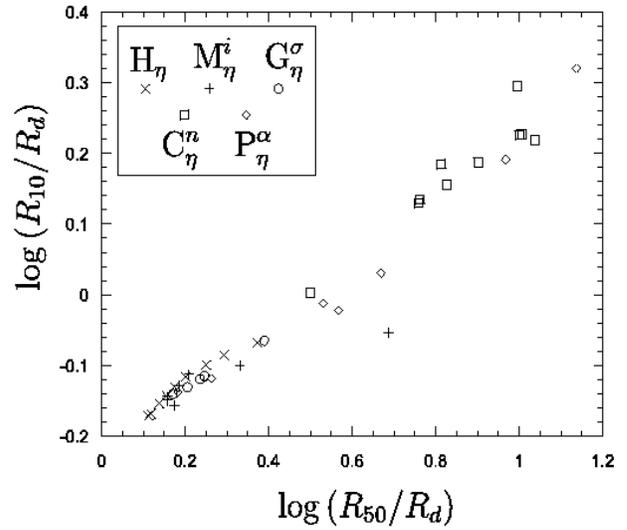}
\end{center}
\caption{The core size segregation: $\ln \left(
R_{10}/r_{c}\right) $ vs $\ln \left( R_{50}/r_{c}\right) $ is plotted for all non
rotating systems.}
\label{segtaille}
\end{figure}

The diagram $\ln \left( R_{10}/R_{d}\right) $ vs $\ln \left(
R_{50}/R_{d}\right) $ is the subject of Figure \ref{segtaille}. One can see
in this figure that each equilibrium state belongs to one or the other
family except in a few particular cases. In the Flat Core family we found
all H$_{\eta }$, G$_{\eta }^{\sigma }$ and M$_{\eta }^{k}$ systems except M$
_{7}^{I}$, and two P$_{\eta }^{\alpha }$ systems namely P$_{50}^{0.5}$ and P$
_{50}^{1}$. These systems are all initially homogeneous or slightly
inhomogeneous (e.g. P$_{50}^{0.5}$ and P$_{50}^{1}$ systems). 
In the Small Core family, we found all C$_{\eta }^{20}$ and the
P$_{50}^{2.0}$ and P$_{09}^{2.0}$ systems.
These systems are all initially rather very inhomogeneous. Finally, there are 5 systems in-between the two categories: C$_{10}^{03}$, P$_{40}^{1.5}$, P$_{08}^{1.5}$, P$_{10}^{1}$ and M$_{07}^{I}$. This last model is the only one which migrates from Flat core set (when $t\simeq 10 Td$) to the edge of the small core region (when $t\simeq 100 Td$).

\subsection{Equilibrium Distribution Function}

\begin{figure}
\begin{center}
\includegraphics[width=84mm]{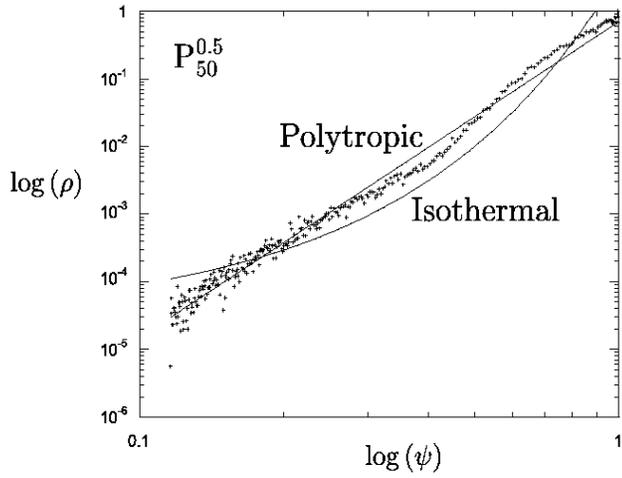}
\end{center}
\caption{Polytropic and isothermal fit for the $P_{50}^{0.5}$ simulation.}
\label{exempfit}
\end{figure}

In order to compare systems in the whole phase space, we fitted the
equilibrium state reached by each system with two distinct isotropic models,
e.g. polytropic and isothermal (see equations \ref{df1}-\ref{df4} or 
BT87, p.223-232). Figure \ref{exempfit} shows these two fits for the P$
_{50}^{0.5}$ simulation. The technique used for the fit is described in
section 2 of this paper. The result obtained for this special study is the
following: the equilibrium states reached by our initial conditions can be
fitted by the two models with a good level of accuracy.
As long as $\eta  <  70$, the polytropic fit gives a mean value $
\gamma =4.77$ with a standard deviation of $\,2.48\;10^{-1}$. This deviation
represents $5.1\%$ of the mean value. This value corresponds typically to the
well known Plummer model for which $\gamma =5$ (see BT87 P.224 for details). When the
collapse is very quiet ( typically $\eta  >  70$ ) polytropic fit is
always very good but the value of the index is much larger than Plummer
model, e.g. $\gamma =6.86$ for H$_{79}$ and $\gamma =7.37$ for H$_{88}$. The
corresponding plot is the subject of the Figure \ref{fitpoly}. All the data can be found
in appendix.
As we can see on the example plotted in Figure \ref{exempfit}, the isothermal
fit is generally not as good as the polytropic one. On the whole set of equilibria,
isothermal fits give a mean value $s^{2}=2.5\,\,10^{-2}$ with a standard
deviation of $1.6\,\,10^{-2}$ ($60\%$). The corresponding plot is the subject
of Figure \ref{fitiso}. All the data can be found in appendix.

\begin{figure}
\begin{center}
\includegraphics[width=84mm]{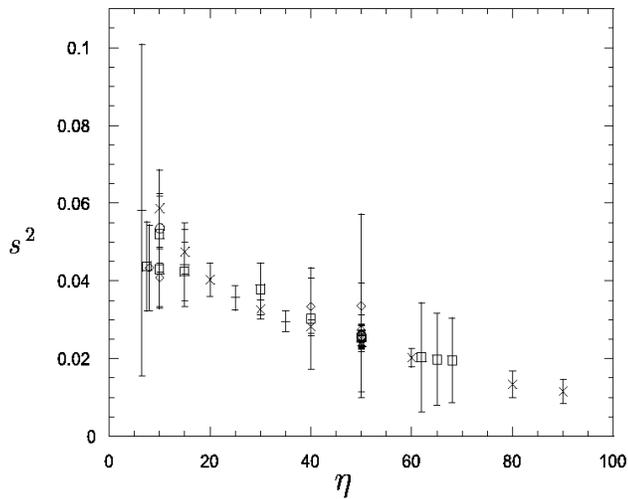}
\end{center}
\caption{Best fit of the $s^{2}$ parameter for an isothermal model for all
non rotating systems studied. The error bar correspond to the least square
difference between the data and the model.}
\label{fitiso}
\end{figure}
\begin{figure}
\begin{center}
\includegraphics[width=84mm]{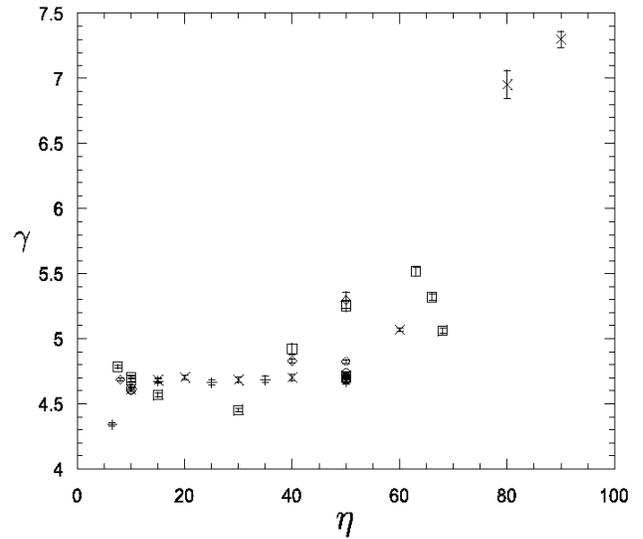}
\end{center}
\caption{Best fit of the $\gamma $ parameter for a polytropic model for all
non rotating systems studied. The error bars correspond to the least square
difference between the data and the model.}
\label{fitpoly}
\end{figure}
In fact both isothermal and polytropic fits are reasonable: as long as the
model is able to reproduce a core halo structure the fit is correct. The
success of the Plummer model, which density is given by 
\[
\rho \left( r\right) =\frac{3}{4\pi b^{3}}\left[ 1+\left( \frac{r}{b}\right)
^{2}\right] ^{-5/2} 
\]
can be explained by its ability to fit a wide range of models with various
ratio of the core size over the half-mass radius. The adjustment of this
ratio is made possible by varying the free parameter $b$. We expect that other core
halo models like King or Hernquist models work as well as the Plummer
model. As a conclusion of this section, let us say that as predicted by
theory there is not a single universal model to describe the equilibrium state of
isotropic spherical self-gravitating system.

\subsection{Influence of rotation}

We saw in section \ref{morphseg} a source of flattening for  self-gravitating equilibrium.
Let us now show the influence of initial rotation, which is a natural candidate to produce flattening.
The way we have added a global rotation and the significance of our
rotation parameters $f$ and $\mu $ are explained in section \ref{secrotation}.
\begin{figure} 
\begin{center}
\includegraphics[width=84mm]{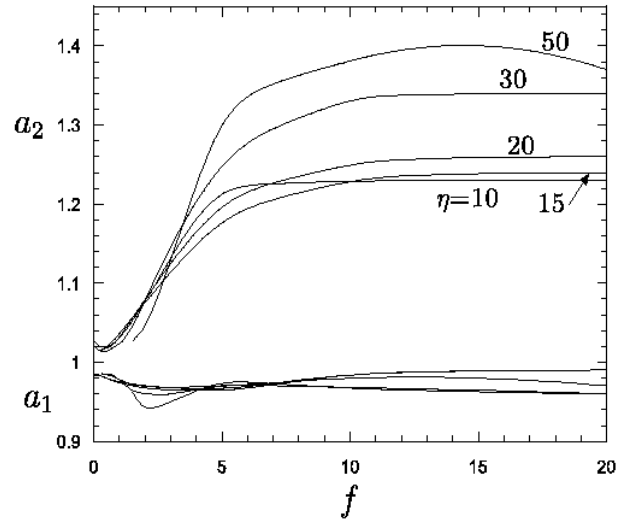}
\end{center}
\caption{Axial ratios as for different values of $\eta$ as a function of the
initial solid rotation parameter $f$}
\label{rotation}
\end{figure}
The set of simulations performed for this study contains 31 different
elements. The initial virial ratio ranges from $\eta =10$ to $\eta =50$, and
the rotation parameter from $f=0$ (i.e. $\mu=0$) to $f=20$ (i.e. $\mu=0.16$ when $\eta = 50$). As a matter of fact,
equilibrium states always preserve a rather important part of the initial rotation\footnote{We observed that $\mu $ is always smaller in the equilibrium state than in the initial one, typically each rotating systems conserves 65\% of the initial $\mu$}
and, observed elliptical gravitating systems generally possess very small amount of rotation (see e.g. \cite{Combes}). 
We thus exclude large values of $f$ . \newline
Our experiments exhibit two main features (see Figure \ref{rotation}): on the one hand, rotation produces a flattened equilibrium state only when $f$ exceeds a triggering value (typically $f=f_o \simeq 4$). On the other hand, we have found that for a given value of $\eta$, the flatness of the equilibrium is roughly $f-$independent, provided that $f>f_o$.

\subsection{Thermodynamical segregation}

\begin{figure}
\begin{center}
\includegraphics[width=84mm]{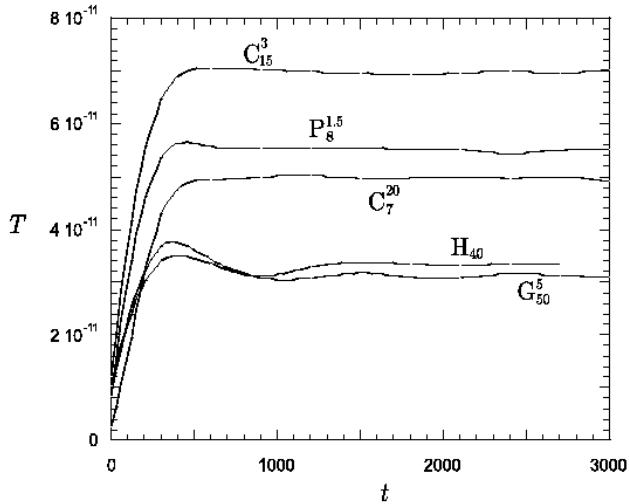}
\end{center}
\caption{Evolution of the temperature as a function of time}
\label{stabtemp}
\end{figure}

\begin{figure}
\begin{center}
\includegraphics[width=84mm]{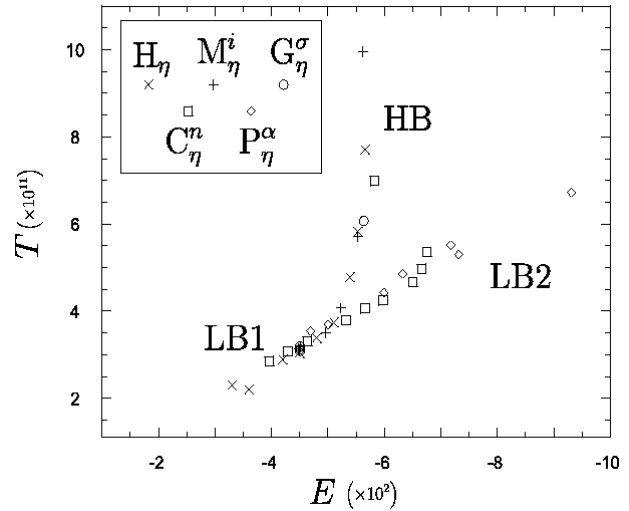}
\end{center}
\caption{Energy-Temperature diagram}
\label{et}
\end{figure}

As we study isolated systems, the total energy $E$ contained in the system
is constant during the considered dynamical evolution . This property
remains true as long as we consider collisionless evolutions. For
gravitational systems, this means that we cannot carry out any simulation of duration
larger than a few hundred dynamical times. We have obviously taken
this constraint into account in our experiments. All systems which experience a violent relaxation reach 
an equilibrium state which is stationary in the
whole phase space. Spatial behaviour like morphological segregation produced
by ROI was confirmed and further detailled thanks to our study. A new size segregation was
found in section \ref{sizeseg}. Now let us consider another new segregation
appearing in the velocity space. Each equilibrium state is associated with a
constant temperature $T$, calculated using equation $\left( \ref{temperature}
\right) $. More precisely, we have calculated the temporal mean value
\footnote{
The temporal mean value is computed from the time when the equilibrium is
reached until the end of the simulation} of the temperatures, evaluated every
one hundred time units. As we can see in Figure \ref{stabtemp}, after the
collapse and whatever the nature of the initial system is, the temperature is a
very stable parameter. \\
Figure \ref{et} shows the $E-T$ diagram of the set of
all non rotating simulations. It reveals a very interesting feature of post-collapse
self-gravitating systems. 

On the one hand, the set of systems with a total energy $E>-0.054$ can
be linearly fitted in the $E-T$ plane. We call this set Low Branch 1
(hereafter denoted by LB1, see Figure \ref{et}). On the other hand, the set of the systems with
a total energy $E<-0.054$ splits into two families. The first is an exact
continuation of LB1. Hence we named it Low Branch 2 (hereafter denoted by LB2). The
second can also be linearly fitted, but with a much greater slope (one
order of magnitude). We label this family High Branch (hereafter denoted by HB).
\newline
In LB1 or LB2, we find every H, G, and M systems with $\eta >25$, every P
and every C with $n>10$. In HB, we find  C$_{10}^{03}$ and every H, G and M systems with $\eta <25$. This segregation thus affects violent collapses (cold initial data): on the one hand, when $
\eta >25$ all systems are on LB1, on the other hand for $\eta <25$,
initially homogeneous or quasi-homogeneous (e.g. C$^{03}_{10}$) systems reach HB when initially inhomogeneous systems
stay on LB2 instead.

\section{Interpretations, Conclusions and Perspectives}\label{intcon}
Let us now recapitulate the results we have obtained and propose an
interpretation:

\begin{enumerate}
\item  The equilibrium state produced by the collapse of a set of $N$
gravitating particles is $N-$independent provided that $N>3.0\,10^{4}$.

\item  Without any rotation, the dissipationless collapse of a set of
gravitating particles can produce two relatively distinct equilibrium states:

\begin{itemize}
\item  If the initial set is homogeneous, the equilibrium has a large core
and a steep envelope.

\item  If the initial set contains significant inhomogeneities ($n >
10$ for clumpy systems or $\alpha  >  1$ for power law systems), the
equilibrium state has only a small core around of which the density falls
down regularly.
\end{itemize}

The explanation of this core size segregation is clear: it is associated to
the Antonov core-collapse instability occurring when the density contrast 
between central and outward region of a gravitating system is very big. As a
matter of fact, if the initial set contains inhomogeneities, these
collapse much more quickly than the whole system\footnote{Because their Jeans length is much more smaller than the one of the whole system.} and fall quickly into the
central regions. The density contrast  becomes then very large and the
Antonov instability triggers producing a core collapse phenomenon. 
The rest of the system then smoothly collapses around this collapsed core.
If there are no inhomogeneities in the initial set, the system collapses
as a whole, central density grows slowly without reaching the triggering value
of the Antonov instability. A large core then forms. Later evolution can
also produce core collapse: this is what occurs for our M$_{07}^{I}$ system (see
Figure \ref{comparplot}). This is an initially homogeneous system with Kroupa mass
spectrum which suffers a very strong collapse. As the mass spectrum is not
sufficient to bring quickly enough a lot of mass in the center of the system,
Antonov instability does not trigger and a large core forms. As the collapse
is very violent, an increasing significant part of particles are
progressively ''ejected'' and the core collapse takes progressively place.
This is the same phenomenon which is generally invoked to explain the collapsed core of some old globular
clusters (e.g. \cite{Djorg}): during its dynamical
evolution in the galaxy, some stars are tidally extracted from a globular
cluster, to compensate this loss the cluster concentrate its core, increasing
then the density contrast, triggering sooner or later the Antonov instability.

\item  Without any rotation, the collapse (violent or quiet) of an
homogeneous set of gravitating particles produces an E0 (i.e. spherical)
isotropic equilibrium state. There are two possible ways to obtain a flattened
equilibrium: 
\begin{itemize}
\item  Introduce a large amount of inhomogeneity  near the center in the initial state, and
make a violent collapse ($\eta <25$).
\item  Introduce a sufficient amount ($f>4$) of rotation in the initial
state.
\end{itemize}
These two ways have not the same origin and do not produce the same equilibrium
state. \newline
In the first case, one can reasonably invoke the Radial Orbit
Instability: as a matter of fact, as it is explained in a lot of works (see 
\cite{PAA98} for example) two features are associated to this phenomenon.
First of all, it is an instability which needs an equilibrium state from
which it grows. Secondly, it triggers only when a sufficient amount of radial
orbits are present.
The only non rotating flattened systems we observed just combine this two conditions:
sufficient amount of radial orbits because the collapse is violent and something from which ROI can 
grow because we have seen in the previous point that inhomogeneities collapse first and quickly join the center. The fact that cold P$_{\eta}^{\alpha}$ systems are more flattened than C$_{\eta}^{\alpha}$ ones is in complete accordance with our interpretation: as a matter of fact, by construction, power law systems have an initial central overdensity, whereas clumpy systems create (quickly but not instantaneously) this overdensity bringing the collapsed clumps near the center.
The ROI flattening is oblate ($a_{2}\simeq 1$ and $a_{1}<1$). \\
The rotational flattening is more natural and occurs when the centrifugal
force overcomes the gravitational pressure. The rotational\ flattening is prolate (
$a_{2}>1$ and $a_{1}\simeq 1$). We notice that initial rotation must be
invoked with parsimony to explain the ellipticity of some globular
clusters or elliptical galaxies.
As a matter of fact, these objects are very weakly rotating systems and our study has shown that the amount of rotation is almost constant during the collapse. 

\item  Spherical equilibria can be suitably fitted by both isothermal and
polytropic laws with various indexes. It suggests that any
distribution function of the energy exhibiting an adaptable core halo
structure (Polytrope, Isothermal, King, Hernquist,...) can suitably fit the equilibrium
produced by the collapse of our initial conditions.

\item  There exists a temperature segregation between equilibrium states. It
concerns only initially cold systems (i.e. systems which will suffer a violent
collapse): for such systems when $\eta $ decreases, the equilibrium
temperature $T$ increases much more for initially homogeneous systems than
for initially inhomogeneous systems. On the other hand, whatever their
initial homogeneity, quiet collapses are rather all equivalent from the point of view of the
equilibrium temperature: $T$ increases in the same way for
all systems as $\eta $ decreases. 
This feature may be the result of the larger influence of the dynamical friction induced 
by the primordial core on the rapid particles in a violent collapse.
\end{enumerate}

All these properties may be directly confronted to physical data from
globular clusters (see Harris catalogue \cite{Harris}) or galaxies observations. 
 \\
As a matter of fact, in the standard "bottom-up" scenario of the hierarchical growth of structures, galaxies naturally form from very inhomogeneous medium. Our study then suggests for the equilibrium state of such objects a potential flattening and a collapsed core. This is in very good accordance with the E0 to E7 observed flatness of elliptical galaxies and may be a good explanation for the presence of massive black hole in the center of galaxies (see \cite{Schodel}). \\
On the other hand Globular Clusters observations show that these are spherical objects (the few low flattened clusters all possess a low amount of rotation), and that their core is generally not collapsed (the collapsed core of almost 10\% of the galactic Globular Clusters can be explained by their dynamical evolution through the galaxy). Our study then expect that Globular Clusters  form from homogeneous media.
\\
These conclusions can be tested using the $E-T$ plane. As a matter of fact, we expect that an $E-T$ plane build from galactic data would not present any High Branch whereas the same plane build from Globular Clusters data would.

\section*{Acknowledgments}
We thank the referee for the relevance of his remarks and suggestions.
We thank Joshua E. Barnes, who wrote the original Treecode. We also
particularly thank Daniel Pfenniger for the use of the parallel Treecode.
The simulations were done on the Beowulf cluster at the Laboratoire de
Math\'ematiques Appliqu\'ees from the \'Ecole Nationale Sup\'erieure de Techniques Avanc\'ees. \\
This is a preprint of an article accepted for publication in {\it Monthly Notices of The Royal Astronomical Society} \copyright 2003 The Royal Astronomical Society.

\appendix
\section{Tables of results}

\begin{table*}
\caption{Homogeneous Initial Conditions: H$_{\eta }$}
\begin{tabular}{c c c c c c c c c c c c c c c c}
$\eta $ &  & $\Delta \;E$ & $\eta $ & $a_{1}$ & $a_{2}$ & $R_{10}$ & $R_{50}$
& $R_{90}$ & $R_{d}$ & $\gamma $ & $\Sigma _{\gamma }^{2}$ & $s^{2}$ & $
\Sigma _{s^{2}}^{2}$ & $T$ & $-E$ \\ 
&  & {\scriptsize ( \%)} & {\scriptsize (end)} &  &  &  &  &  &  &  & 
{\scriptsize (}$\times ${\scriptsize \ 10}$^{2}${\scriptsize )} & 
{\scriptsize (}$\times ${\scriptsize \ 10}$^{2}${\scriptsize )} & 
{\scriptsize (}$\times ${\scriptsize \ 10}$^{3}${\scriptsize )} & 
{\scriptsize (}$\times ${\scriptsize \ 10}$^{11}${\scriptsize )} & 
{\scriptsize (}$\times ${\scriptsize \ 10}$^{2}${\scriptsize )} \\ 
\hline\hline
88 &  &  0.0 &  98 &  1.02 &  0.99 &  3.39 &  6.53 &  10.6 &  5.02 & 7.37 & -6 & 1.12 & -3 &  2.30 &  3.30 \\ 
79 &  &  0.0 &  99 &  1.00 &  1.00 &  3.11 &  6.01 &  9.8 &  4.59 &  6.86  &  -4 &  1.37 &  -3 &  2.20 & 3.60 \\ 
60 &  &  0.0 &  96 &  1.01 &  0.98 &  2.41 &  4.73 &  11.5 &  3.41 & 5.05 &  -1 &  2.04 &  -2 &  2.89 & 4.20 \\ 
50 &  &  0.0 &  96 &  1.01 &  0.98 &  2.04 &  4.09 &  15.1 &  2.81 & 4.73 &  -2 &  2.49 &  -2 &  3.04 & 4.50 \\ 
40 &  &  0.1 &  96 &  1.01 &  0.99 &  1.72 &  3.51 &  25.7 &  2.30 & 4.72 &  -2 &  2.79 &  -2 &  3.39 & 4.80 \\ 
30 &  &  0.0 &  96 &  1.02 &  0.99 &  1.36 &  2.88 &  253.2 &  1.75 & 4.66 &  -2 &  3.27 &  -3 &  3.74 & 5.10 \\ 
20 &  &  0.0 &  101 &  1.01 &  0.99 &  0.95 &  2.22 &  874.1 &  1.17 & 4.68 &  -1 &  4.03 &  -5 &  4.78 & 5.39 \\ 
15 &  &  0.0 &  108 &  1.01 &  0.99 &  0.74 &  1.89 &  1143.0 &  0.88 & 4.66 &  -1 &  4.73 &  -6 &  5.83 & 5.53 \\ 
10 &  &  1.4 &  120 &  1.02 &  0.98 &  0.52 &  1.59 &  1448.0 &  0.60 & 4.59 &  -1 &  5.76 &  -9 &  7.71 & 5.66
\end{tabular}
\end{table*}

\begin{table*}
\caption{Clumpy Initial Condition: C$_{\eta }^{n}$}
\begin{tabular}{c c c c c c c c c c c c c c c c}
$\eta $ & $n$ & $\Delta \;E$ & $\eta $ & $a_{1}$ & $a_{2}$ & $R_{10}$ & $
R_{50}$ & $R_{90}$ & $R_{d}$ & $\gamma $ & $\Sigma _{\gamma }^{2}$ & $s^{2}$
& $\Sigma _{s^{2}}$ & $T$ & $-E$ \\ 
&  & {\scriptsize ( \%)} & {\scriptsize (end)} &  &  &  &  &  &  &  & 
{\scriptsize (}$\times ${\scriptsize \ 10}$^{2}${\scriptsize )} & 
{\scriptsize (}$\times ${\scriptsize \ 10}$^{2}${\scriptsize )} & 
{\scriptsize (}$\times ${\scriptsize \ 10}$^{3}${\scriptsize )} & 
{\scriptsize (}$\times ${\scriptsize \ 10}$^{11}${\scriptsize )} & 
{\scriptsize (}$\times ${\scriptsize \ 10}$^{2}${\scriptsize )} \\ 
\hline\hline
 10 &  3 &  0.1 &  97 &  1.03 & 0.96 &  0.55 &  1.85 &  1241.0 & 0.58 &  4.61 &  -1 &  5.21 &  -11 & 6.99 &  5.82 \\ 
67 &  20 &  0.8 &  95 &  1.01 & 0.94 &  0.93 &  6.14 &  16.1 &  0.66 &  5.00 &  -1 &  1.98 &  -10 &  2.86 &  3.96 \\ 
65 &  20 &  0.9 &  95 &  1.02 & 0.96 &  0.93 &  5.63 &  13.5 &  0.63 &  5.27 &  -2 &  2.00 &  -13 &  3.07 &  4.29 \\ 
61 &  20 &  0.6 &  95 &  1.05 & 0.98 &  0.86 &  5.15 &  12.3 &  0.58 &  5.43 &  -3 &  2.09 &  -11 &  3.32 &  4.64 \\ 
48 &  20 &  0.6 &  94 &  1.03 & 0.99 &  0.81 &  4.24 &  11.8 &  0.59 &  5.15 &  -3 &  2.61 &  -13 &  3.79 &  5.31 \\ 
39 &  20 &  0.5 &  94 &  1.04 & 0.99 &  0.76 &  3.84 &  12.8 &  0.40 &  4.72 &  -1 &  3.20 &  -10 &  4.06 &  5.65 \\ 
29 &  20 &  1.2 &  94 &  1.04 & 0.99 &  0.72 &  3.42 &  15.4 &  0.56 &  4.42 &  -1 &  3.81 &  -6 &  4.24 &  5.97 \\ 
14 &  20 &  0.3 &  93 &  1.09 & 0.98 &  0.64 &  2.72 &  39.5 &  0.48 &  4.56 &  -1 &  4.27 &  -7 &  4.66 &  6.50 \\ 
10 &  20 &  1.4 &  97 &  1.13 & 0.99 &  0.61 &  2.54 &  345.5 & 0.54  &  4.70 &  -9 &  4.25 &  -15 & 4.98  &  6.66 \\ 
7 &  20 &  0.3 &  94 &  1.14 & 0.92 &  0.57 &  2.44 &  224.6 & 0.45  &  4.74 &  -1 &  4.41 &  -12 & 5.36  &  6.76
\end{tabular}
\end{table*}

\begin{table*}
\caption{Power Law Initial Conditions: P$_{\eta }^{\alpha }$}
\begin{tabular}{c c c c c c c c c c c c c c c c}
$\eta $ & $\alpha $ & $\Delta \;E$ & $\eta $ & $a_{1}$ & $a_{2}$ & $R_{10}$
& $R_{50}$ & $R_{90}$ & $R_{d}$ & $\gamma $ & $\Sigma _{\gamma }^{2}$ & $
s^{2}$ & $\Sigma _{s^{2}}$ & $T$ & $-E$ \\ 
&  & {\scriptsize ( \%)} & {\scriptsize (end)} &  &  &  &  &  &  &  & 
{\scriptsize (}$\times ${\scriptsize \ 10}$^{2}${\scriptsize )} & 
{\scriptsize (}$\times ${\scriptsize \ 10}$^{2}${\scriptsize )} & 
{\scriptsize (}$\times ${\scriptsize \ 10}$^{3}${\scriptsize )} & 
{\scriptsize (}$\times ${\scriptsize \ 10}$^{11}${\scriptsize )} & 
{\scriptsize (}$\times ${\scriptsize \ 10}$^{2}${\scriptsize )} \\ 
\hline\hline
50 &  0.5 &  0.0 &  95 &  1.01 & 0.99 &  1.84 &  3.92 &  13.5 &  2.53 &  4.66 &  -1 &  2.65 &  -2 &  3.54 &  4.69 \\ 
50 &  1 &  0.0 &  94 &  1.01 & 0.99 &  1.56 &  3.77 &  12.1 &  2.01 &  4.77 &  -6 &  2.78 &  -3 &  3.69 &  5.00 \\ 
10 &  1 &  0.1 &  96 &  1.00 & 0.80 &  0.69 &  2.71 &  382.2 & 0.70  &  4.61 &  -8 &  4.05 &  -8 & 4.85 &  6.32 \\ 
8 &  1.5 &  0.1 &  96 &  1.01 & 0.71 &  0.62 &  2.63 &  25.1 &  0.61 &  4.63 &  -7 &  4.42 &  -9 &  5.52 &  7.18 \\ 
50 &  2 &  1.7 &  93 &  1.02 & 0.99 &  0.53 &  3.20 &  9.2 &  0.34 &  5.30 &  -6 &  3.35 &  -9 &  5.30 &  7.32 \\ 
40 &  1.5 &  0.1 &  96 &  1.00 & 0.99 &  0.97 &  3.31 &  11.0 &  1.03 &  4.71 &  -8 &  3.44 &  -6 &  4.42 &  5.99 \\ 
9 &  2 &  1.6 &  96 &  1.01 & 0.78 &  0.38 &  2.51 &  10.6 &  0.18 &  4.68 &  -10 &  5.21 &  -20 &  6.73 &  9.31
\end{tabular}
\end{table*}

\begin{table*}
\caption{Mass Spectrum Initial Conditions: M$_{\eta }^{i}$}
\begin{tabular}{ c c c c c c c c c c c c c c c c }
$\eta $ & i & $\Delta \;E$ & $\eta $ & $a_{1}$ & $a_{2}$ & $R_{10}$ & $
R_{50} $ & $R_{90}$ & $R_{d}$ & $\gamma $ & $\Sigma _{\gamma }^{2}$ & $s^{2}$
& $\Sigma _{s^{2}}$ & $T$ & $-E$ \\ 
&  & {\scriptsize (\% )} & {\scriptsize (end)} &  &  &  &  &  &  &  & 
{\scriptsize (}$\times ${\scriptsize \ 10}$^{2}${\scriptsize )} & 
{\scriptsize (}$\times ${\scriptsize \ 10}$^{2}${\scriptsize )} & 
{\scriptsize (}$\times ${\scriptsize \ 10}$^{3}${\scriptsize )} & 
{\scriptsize (}$\times ${\scriptsize \ 10}$^{11}${\scriptsize )} & 
{\scriptsize (}$\times ${\scriptsize \ 10}$^{2}${\scriptsize )} \\ 
\hline\hline
7 &  Krou &  5.0 &  132 &  1.02 & 0.99 &  0.25 &  2.04 &  1721.0 & 0.37 &  4.26 &  -1 & 5.16 &  -15 & 9.97 &  5.62 \\ 
15 &  1/M &  0.6 &  101 &  1.01 & 0.98 &  0.68 &  2.04 &  366.8 & 0.97  &  4.65 &  -1 & 4.35 &  -10 & 5.72  &  5.53 \\ 
25 &  Salp &  0.4 &  99 &  1.01 & 0.98 &  1.18 &  2.51 &  225.5 & 1.54  &  4.66 &  -2 & 3.59 &  -3 & 4.08 &  5.23 \\ 
35 &  Krou &  0.2 &  98 &  1.01 & 0.99 &  1.55 &  3.18 &  39.8 &  2.09 &  4.66 &  -2 &  2.97 &  -3 &  3.51 &  4.95 \\ 
51 &  1/M &  0.2 &  95 &  1.01 & 0.98 &  1.79 &  4.19 &  14.2 &  2.83 &  4.67 &  -9 &  2.39 &  -4 &  3.15 &  4.48 \\ 
50 &  Krou &  0.1 &  96 &  1.02 & 0.98 &  1.93 &  4.09 &  15.1 &  2.87 &  4.60 &  -1 &  2.45 &  -3 &  3.19 &  4.50 \\ 
50 &  Salp &  0.1 &  96 &  1.01 & 0.98 &  2.03 &  4.13 &  15.1 &  2.87 &  4.70 &  -2 &  2.45 &  -2 &  3.08 &  4.49
\end{tabular}
\end{table*}

\begin{table*}
\caption{Gaussian Velocity Dispersion Initial Conditions: G$_{\eta }^{\sigma }$}
\begin{tabular}{ c c c c c c c c c c c c c c c c }
$\eta $ & $\sigma $ & $\Delta \;E$ & $\eta $ & $a_{1}$ & $a_{2}$ & $R_{10}$
& $R_{50}$ & $R_{90}$ & $R_{d}$ & $\gamma $ & $\Sigma _{\gamma }^{2}$ & $
s^{2}$ & $\Sigma _{s^{2}}^{2}$ & $T$ & $-E$ \\ 
&  & {\scriptsize (\% )} & {\scriptsize (end)} &  &  &  &  &  &  &  & 
{\scriptsize (}$\times ${\scriptsize \ 10}$^{2}${\scriptsize )} & 
{\scriptsize (}$\times ${\scriptsize \ 10}$^{2}${\scriptsize )} & 
{\scriptsize (}$\times ${\scriptsize \ 10}$^{3}${\scriptsize )} & 
{\scriptsize (}$\times ${\scriptsize \ 10}$^{11}${\scriptsize )} & 
{\scriptsize (}$\times ${\scriptsize \ 10}$^{2}${\scriptsize )} \\ 
\hline\hline
48 &  G1 &  0.0 &  95 &  1.00 & 1.00 &  1.72 &  3.98 &  14.6 &  2.23 &  4.66 &  -5 &  2.64 &  -3 &  3.13 &  4.50 \\ 
49 &  G2 &  0.0 &  95 &  1.02 & 0.99 &  1.74 &  4.02 &  14.4 &  2.28 &  4.65 &  -6 &  2.61 &  -3 &  3.13 &  4.50 \\
50 &  G3 &  0.0 &  95 &  1.00 & 1.00 &  1.90 &  4.08 &  14.1 &  2.56 &  4.72 &  -8 &  2.52 &  -2 &  3.09 &  4.50 \\ 
12 &  G4 &  0.4 &  118 &  1.03 & 0.98 &  0.56 &  1.77 &  1312.0 & 0.64 &  4.56 &  -1 &  5.33 &  -10 & 6.08 &  5.63 \\ 
50 &  G5 &  0.0 &  96 &  1.01 & 0.99 &  1.98 &  4.11 &  14.8 &  2.72 & 4.71 &  -1 &  2.50 &  -2 & 3.19 & 4.50
\end{tabular}
\end{table*}

\end{document}